\documentclass{PoS}
\usepackage{subcaption}
\usepackage{lineno}

\title{Results from the first one and a half years of the HAWC GRB program}

\ShortTitle{Results from 1.5 years of the HAWC GRB program}

\author{\speaker{Joshua Wood} $^a$ for the HAWC Collaboration$^b$\\
        \llap{$^a$}Department of Physics, University of Wisconsin, Madison, Wisconsin, USA \\
        \llap{$^b$}For a complete author list, see the special section of these proceedings. \\
        E-mail: \email{jwood@icecube.wisc.edu}}

\abstract{The High-Altitude Water Cherenkov (HAWC) Observatory is a ground-based
TeV gamma-ray observatory in the state of Puebla, Mexico at an altitude of 4100 m
above sea level. Its $\sim$22,000 m$^2$ instrumented area, wide field of view (2 sr),
and $>$95\% uptime make it an ideal instrument for discovering gamma-ray burst (GRB)
emission at $>$100 GeV. Such a discovery would provide key information about the
origins of prompt GRB emission as well as constraints on EBL models and the violation
of Lorentz invariance. We present here the results of our current GRB search methods,
which include an all-sky search as well as fast follow-ups of GRBs reported by satellites,
after one and a half years of data with the full HAWC detector.}

\FullConference{35th International Cosmic Ray Conference - ICRC2017 -\\10-20 July, 2017\\Bexco, Busan, Korea}

\begin{document}

\section{Introduction}

The HAWC Observatory is a ground-based TeV gamma-ray observatory located in the state
of Puebla, Mexico at an altitude of 4100 m above sea level. Its $\sim$22,000 m$^2$
instrumented area, wide field of view (2 sr), and $>$95\% uptime allow it to
continuously monitor for Northern hemisphere transients with emission extending
up to 100 GeV and beyond. This is particularly important for detecting
gamma-ray burst (GRB) transients which have been measured up to 95 GeV
but typically have prompt emission durations shorter than the slewing times of
imaging atmospheric Cherenkov telescopes.

Measurements of the highest energy gamma-rays associated with GRBs are key to developing models of the relativistic
jets powering emission. This is because observations of a spectral cutoff at the highest photon energies can be interpreted as
estimates of the bulk Lorentz factor $\Gamma$ in the region where gamma-rays are produced \cite{Piran:1999kx}, providing
insight into the internal GRB environment as well as the expected neutrino flux from GRBs which
is sensitive to $\Gamma$ \cite{Waxman:1997ti}. Alternatively, interpreting the spectral cutoff as attenuation of GRB photons
from pair-production on extra-galactic background light (EBL) provides constraints on EBL density over cosmological distances \cite{Gilmore:2009zb}.

\section{HAWC GRB Program Overview}

The HAWC GRB program presently consists of two dedicated analyses, 
a self-triggered all-sky search and rapid response follow-ups of GRBs reported
by satellites. Both methods are performed in real-time using the quick-look air shower
reconstruction produced at the HAWC site with a latency of $\sim$5 seconds. Additionally,
each search is repeated on archival data when improved calibrations and reconstruction
algorithms become available.

The self-triggered all-sky method continuously searches for GRB
transients at energies >100 GeV with three sliding time windows of lengths 0.2, 1, and 10 seconds which are typical
of peak structures within GRB light curves. We perform the search by shifting each window forward in time by
10\% its width and binning air shower events during that window using a grid of 2.1$^{\circ}$ $\times$ 2.1$^{\circ}$ square spatial bins
covering all points within 50$^\circ$ of detector zenith \cite{Wood:2016}. We then compare the number of showers in each spatial bin to 
the expectation from charged cosmic-ray backgrounds.
Locations with an air shower excess corresponding to a post-trials false alarm rate of 1 event per day are considered candidates for 
GRB transients and reported internally within HAWC.

The method for rapid response follow-up of reported GRBs is simpler in that it fixes the search window start time to match
the external trigger time provided by a satellite. In addition, the spatial portion of the search is restricted to the reported error
on the GRB location. However the basic idea is the same with air shower events during the search window binned into spatial bins 
and compared against the expectation from charged cosmic-ray backgrounds \cite{Dirk:2017}. In cases when $T_{90}$ is available with the external trigger we
search for emission occurring within $T_{90}$ as well as 3$\times T_{90}$, 10$\times T_{90}$ for long GRBs and 6, 20 seconds for short GRBs
with the longer windows covering possible extended emission. Otherwise we perform the follow-up with timescales of 1 and 20 seconds
to cover typical $T_{90}$ values as well as a 300 second window to look for extended emission.

From the sensitivity of the HAWC Observatory we expect $\sim$0.5 GRB detections per year from following
satellite reported GRBs alone assuming most GRBs do not contain an intrinsic cutoff below 200 GeV \cite{Taboada:2013uza}.
The all-sky search method yields a similar expectation as the trials penalty taken when searching
the full sky is roughly compensated by the ability to search data without satellite coverage \cite{Wood:2016}.
Thus far there have been no significant detections of GRB transients in the first
1.5 years of data taking since the inauguration of the full HAWC detector on March 19, 2015.
This is still consistent with our expectations.

\section{Results}

\begin{figure} [ht!]
  \centering
  \includegraphics[width=6in]{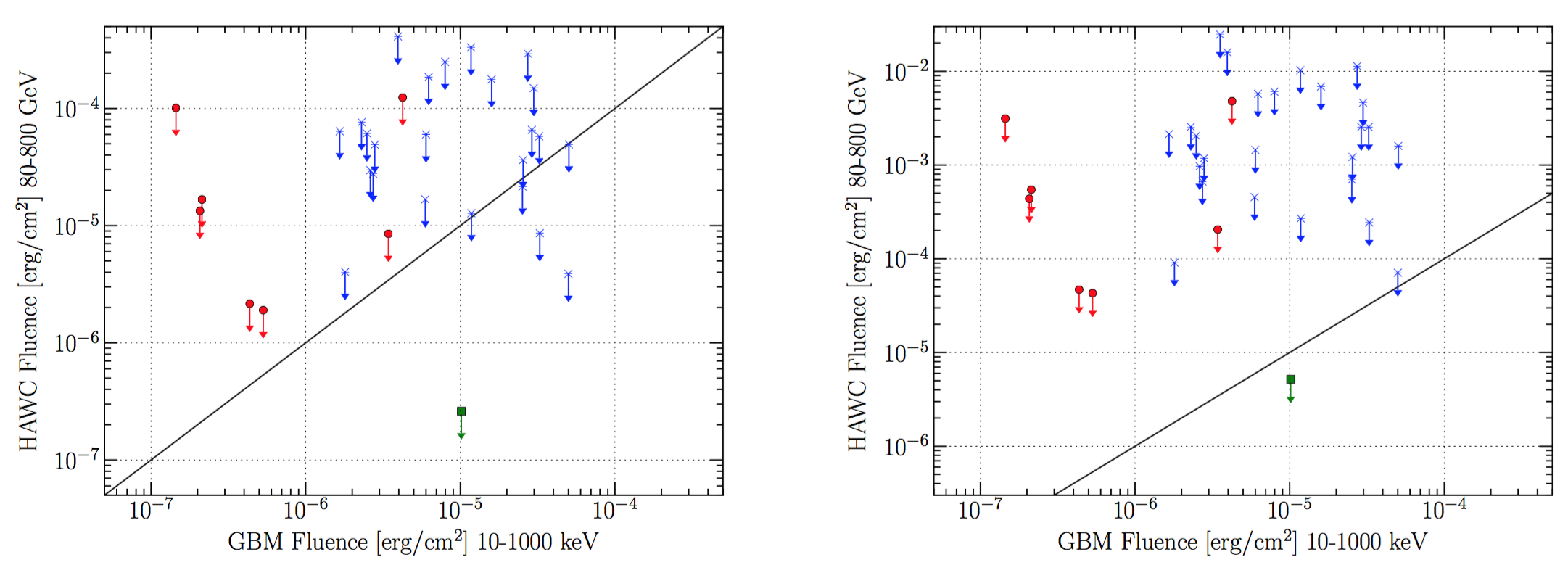}
  \caption
  {Comparison between the Fermi-GBM fluence and the fluence implied by the HAWC
   upper limits obtained during the same time period for all GRBs completely inside
   the HAWC field of view for the different redshifts (left z = 0.3, right z = 1.0).
   The circles (red) show short GRBs, the asterisks (blue) long GRBs and the green
   square GRB 170206A. The black line shows an equal fluence in the Fermi-GBM and HAWC energy range.
   See reference \cite{Dirk:2017} for the complete list of GRBs contributing to these plots.
  }
  \label{fig:triggered_grb_limits}
\end{figure}

\begin{figure} [ht!]
  \centering
  \includegraphics[width=3.5in]{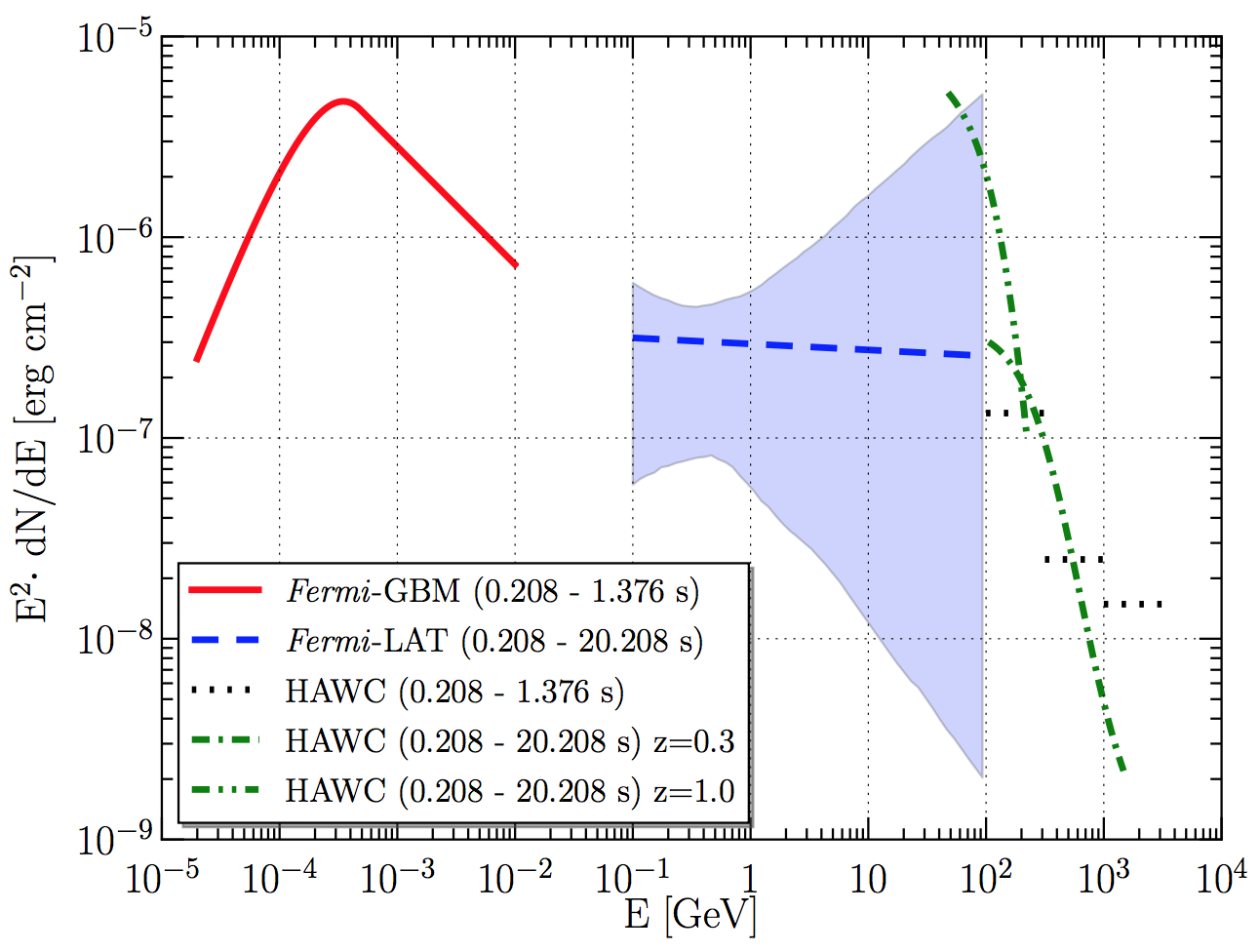}
  \caption
  {Solid red line shows the spectrum fitted to the prompt Fermi-GBM data of
   GRB 170206A, while dotted black lines show the quasi-differential HAWC limits
   assuming an $E^{-2}$ spectrum obtained from the HAWC data taken during the same time period.
   The dashed blue line shows the best fit spectrum obtained from the Fermi-LAT data
   in the early afterglow and the shaded area the uncertainty taking into account
   correlations and non-linearity of fit parameters. The green dashed- dotted lines
   show the HAWC limits for two different assumed redshifts.
  }
  \label{fig:triggered_grb_170206_limit}
\end{figure}

Given the absence of a detection we have placed upper limits on very high energy (VHE) emission
above 80 GeV occurring within the reported $T_{90}$ of 64 GRBs detected by satellites within the HAWC field of
view during the first 1.5 years with the full detector (Figure \ref{fig:triggered_grb_limits}).
This includes a limit on the third brightest burst seen by the Fermi Gamma-ray
Burst Monitor, GRB 170206A, which was also detected by the Fermi Large Area Telescope (Figure \ref{fig:triggered_grb_170206_limit}).
We derive these limits assuming an $E^{-2}$ power-law spectrum for each GRB attenuated 
according to EBL absorption with the fiducial model by Gilmore et al. \cite{Gilmore:2009zb} for two different redshifts, $z = 0.3, 1.0$.
See reference \cite{Dirk:2017} for more details. 

\vspace{2cm}

\vspace{1cm}

\newpage

\begin{figure} [ht!]
  \begin{subfigure}{3in}
  \includegraphics[width=2.5in]{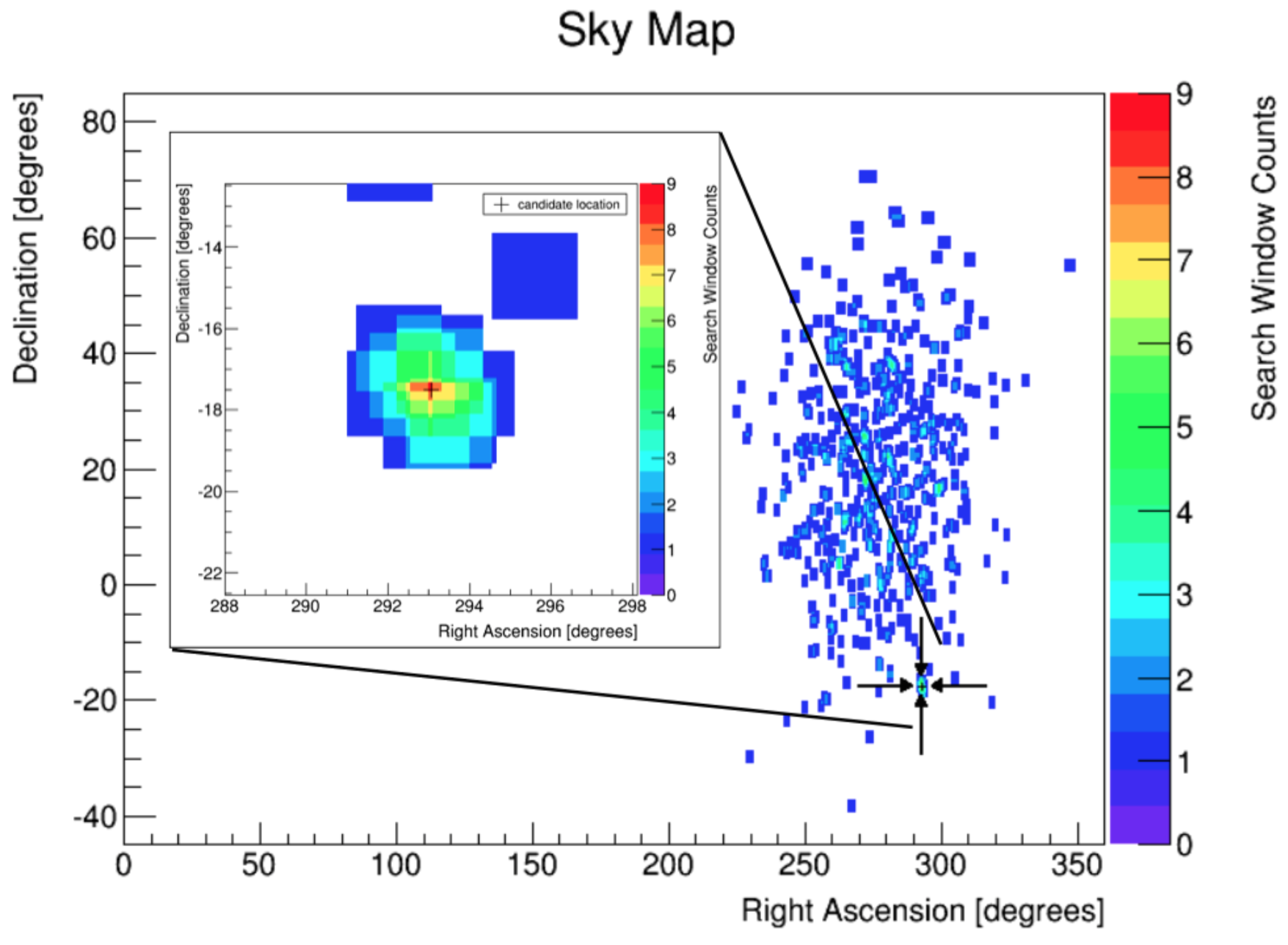}
  \end{subfigure}
  \begin{subfigure}{3in}
  \includegraphics[width=2.8in]{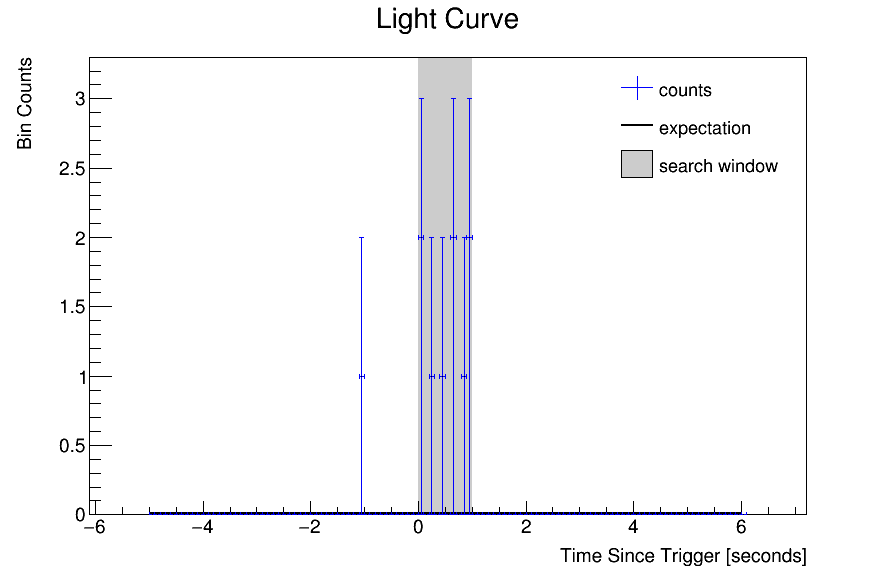}
  \end{subfigure}
  \caption
  {(Left) Sky map of detected air shower counts for the 1 second long time window
   containing the best candidate event from the all-sky GRB search.
   The cross marking the location of the candidate is at a 
   detector zenith angle of 40.5$^\circ$. (Right)
   Light curve of event counts binned in intervals of 0.1 seconds for
   the location of the best candidate. The 1 second window containing
   the best candidate is shaded in grey. The black line marks the background expectation in each light curve bin.
   The background expectation is so low that it cannot be distinguished from zero in this plot.
  }
  \label{fig:best_candidate_plots}
\end{figure}

The best GRB candidate found in the all-sky search method during the first 1.5 years of operating
the full HAWC detector is a 1 second transient with a pre-trial probability of 9$\times10^{-15}$.
Table \ref{tab:1seccand} presents the details of this candidate and Figure \ref{fig:best_candidate_plots}
shows its light curve and sky map. Accounting for the effective trials taken when searching the full sky over this time period with
three search window durations of lengths 0.2, 1, and 10 seconds results in a post-trial probability
of 0.19. Although this result is not significant, it does show the all-sky search algorithm can successfully
find interesting event excesses within the HAWC dataset.

\vspace{1cm}

\begin{table}[ht!]
\begin{center}
\begin{tabular}{|l|c|} \hline
 \multicolumn{2}{|c|}{\textbf{Best All-sky GRB Search Candidate}} \\ \hline
 Date & 2015/05/30 \\
 Trigger Time & 08:20:59.67 UTC \\
 Duration & 1.0 second \\
 Obs. Counts & 9 \\
 Bkg. Counts & 0.115 \\
 Right Ascension & 292.83$^\circ$ (J2000) \\
 Declination & -17.53$^\circ$ (J2000) \\
 Zenith & 40.48$^\circ$ \\ \hline
\end{tabular}
\caption{Details of the overall best candidate from the all-sky GRB search with 1.5 years of data.}
\label{tab:1seccand}
\end{center}
\end{table}


\newpage

\section{Testing for Delayed VHE Emission}

We recently began expanding our follow-up program of GRBs reported by satellites to test for 
delayed VHE emission which is a well known feature of bursts detected by the Fermi Large Area Telescope \cite{Ackermann:2013}.
This was initially done by repeating our standard follow-up analysis with additional time windows starting in integer multiples
of $T_{90}$ after the reported trigger time \cite{Simone:2017}. Additionally, we implemented
the ability to pass external triggers to our all-sky analysis method in order to perform a limited sliding
time window search near the trigger time and location of known GRBs which allows us to finely optimize for
the start time of VHE emission when performing follow-ups.

We tested the new follow-up capability of our all-sky search by re-analyzing several of the same GRBs studied
with our existing GRB follow-up analysis. This was done by setting the sliding time window duration equal to the $T_{90}$
of each burst and reducing the temporal search range to select the single $T_{90}$ window starting at the
externally reported trigger time. Table \ref{tab:comparison} presents the results which are largely identical 
to the standard follow-up analysis. The minor differences in the number of
on-source counts and background expectation arises from the use of square spatial bins in the all-sky search and circular
spatial bins in the standard follow-up analysis.

\newpage

\begin{table}[ht!]
\begin{center}
\begin{tabular}{|c|l|l|} \hline
 \textbf{GRB} & \textbf{Standard follow-up analysis} & \textbf{All-sky analysis with external trigger} \\ \hline
 160605A & n = $\,\,\,\,$4,\quad bg = $\,\,\,\,$0.70,\quad P = 5.8$\times$10$^{-3}$ & n = $\,\,\,\,$4,\quad bg = 0.59 $\pm$ 0.02,\quad P = 3.1$\times$10$^{-3}$ \\
 160410A & n = $\,\,\,\,$2,\quad bg = $\,\,\,\,$2.42,\quad P = 0.70 & n = $\,\,\,\,$2,\quad bg = 2.49 $\pm$ 0.06,\quad P = 0.71 \\
 160310A & n = $\,\,\,\,$6,\quad bg = $\,\,\,\,$4.48,\quad P = 0.29 & n = $\,\,\,\,$5,\quad bg = 4.35 $\pm$ 0.11,\quad P = 0.44 \\
 151228B & n = 36,\quad bg = 39.05,\quad P = 0.71 & n = 38,\quad bg = 38.6 $\pm$ 0.30,\quad P = 0.56 \\
 151205A & n = 36,\quad bg = 32.51,\quad P = 0.29 & n = 34,\quad bg = 30.8 $\pm$ 0.30,\quad P = 0.31 \\ \hline
\end{tabular}
\caption{Comparison of standard follow-up analysis to triggered extension of all-sky analysis for 5 selected GRBs from \cite{Dirk:2017}.
$n$ represents the number of air shower events arriving within $T_{90}$ inside the spatial bin at the GRB location, $bg$ is the
cosmic-ray background for this spatial bin, and $P$ is the Poisson probability for observing $n$ counts given a mean of $bg$.}
\label{tab:comparison}
\end{center}
\end{table}



\section{Summary}

We continue to support and expand upon our existing GRB program.
Although there were no GRB detections within the first 1.5 years of HAWC data we have begun
to place upper limits on emission $>$100 GeV for satellite detected GRBs within the HAWC field of view
with the most constraining limits coming from GRB 170206A.
The latest addition to our GRB program is the ability to test for delayed onset VHE emission
when performing follow-ups of satellite detected GRBs.

\section*{Acknowledgements}
We acknowledge the support from: the US National Science Foundation (NSF); the
US Department of Energy Office of High-Energy Physics; the Laboratory Directed
Research and Development (LDRD) program of Los Alamos National Laboratory;
Consejo Nacional de Ciencia y Tecnolog\'{\i}a (CONACyT), M{\'e}xico (grants
271051, 232656, 260378, 179588, 239762, 254964, 271737, 258865, 243290,
132197), Laboratorio Nacional HAWC de rayos gamma; L'OREAL Fellowship for
Women in Science 2014; Red HAWC, M{\'e}xico; DGAPA-UNAM (grants IG100317,
IN111315, IN111716-3, IA102715, 109916, IA102917); VIEP-BUAP; PIFI 2012, 2013,
PROFOCIE 2014, 2015;the University of Wisconsin Alumni Research Foundation;
the Institute of Geophysics, Planetary Physics, and Signatures at Los Alamos
National Laboratory; Polish Science Centre grant DEC-2014/13/B/ST9/945;
Coordinaci{\'o}n de la Investigaci{\'o}n Cient\'{\i}fica de la Universidad
Michoacana. Thanks to Luciano D\'{\i}az and Eduardo Murrieta for technical support.

\newpage


\providecommand{\href}[2]{#2}\begingroup\raggedright\endgroup

\end{document}